\documentclass[twocolumn,twocolappendix]{aastex63}

\newcommand\ds{{\sc Dark Sage}}
\newcommand\HI{H\,{\sc i}}
\newcommand{\diff}{{\rm d}}

\received{Feb 1, 2022}
\revised{May 10, 2022}
\accepted{\today}
\submitjournal{ApJ}

\def\be{\begin{equation}}
\def\ee{\end{equation}}
\def\CF3{{\sc cosmicflows-3}}
\usepackage{rotating}
\usepackage{multirow}
\usepackage{CJKutf8}
\usepackage{graphicx}	
\usepackage{amsmath}	
\usepackage{amssymb}	
\usepackage{hyperref}
\usepackage{newtxtext,newtxmath}

\usepackage[T1]{fontenc}
\usepackage{ae,aecompl}

\def\Q22{ \href{https://iopscience.iop.org/article/10.3847/1538-4357/ac8b6f}{Q22}  }
 
\usepackage{color}

\shorttitle{Halo Profile for Galaxies}
\shortauthors{Qin et al.}
\graphicspath{{./}{figures/}}

\begin{document}

\title{The Galaxy Number Density Profile of Haloes}

\correspondingauthor{Fei Qin}
\email{feiqin@kasi.re.kr}

\author[0000-0001-7950-7864]{Fei Qin}
\altaffiliation{Korea Astronomy and Space Science Institute, Yuseong-gu, Daedeok-daero 776, Daejeon 34055, Republic of Korea}
\affiliation{Korea Astronomy and Space Science Institute, Yuseong-gu, Daedeok-daero 776, Daejeon 34055, Republic of Korea}
\affiliation{School of Physics, Korea Institute for Advanced Study, Dongdaemun-gu,  Hoegiro 85, Seoul 02455, Republic of Korea}

\author{David Parkinson}
\affiliation{Korea Astronomy and Space Science Institute, Yuseong-gu, Daedeok-daero 776, Daejeon 34055, Republic of Korea}

\author{Adam R.~H.~Stevens}
\affiliation{International Centre for Radio Astronomy Research, The University of Western Australia, Crawley, WA 6009, Australia}

\author{Cullan Howlett}
\affiliation{School of Mathematics and Physics, The University of Queensland, Brisbane, QLD 4072, Australia}



\begin{abstract}

More precise measurements of galaxy clustering will be provided by the next generation of galaxy surveys such as DESI, WALLABY and SKA.  To utilize this information to improve our understanding of the Universe, we need to accurately model the distribution of galaxies in their host dark matter halos. 
In this work we present a new   galaxy number density profile of haloes, which makes predictions for the positions of galaxies in the host halo, different to the widely adopted Navarro–Frenk–White (NFW) profile, since galaxies tend to be found more in the outskirts of halos (nearer the virial radius) than an NFW profile. 
The parameterised   galaxy number density profile model of haloes is fit and tested using the \ds~semi-analytic model of galaxy formation. We find that our   galaxy number density profile model of haloes can accurately reproduce the halo occupation distribution and galaxy two-point correlation function of the \ds~simulation. We also derive the analytic expressions for the circular velocity and gravitational potential energy for this profile model. We use the SDSS DR10 galaxy group catalogue to validate this   galaxy number density profile model of haloes. Compared to the  NFW profile, we find that our model more accurately predicts the positions of galaxies in their host halo and the galaxy two-point correlation function.  

\end{abstract}



\section{Introduction} \label{sec:intro}

Many galaxies tend to gather in groups/clusters of several members, and a group of galaxies will 
inhabit a dark matter halo, for which the dark matter dominates the mass of this system. 
In a group, the most massive or luminous galaxy is identified as the central galaxy, which typically resides in the center of the host halo gravitational potential \citep{Yang2013,Wetzel2013,RanLi2014}, while the rest of the galaxies are labelled `satellites'. 
Studying the galaxy--halo connection and   density profile of halos enables us to understand the formation and evolution of galaxies and constrain cosmological models \citep{White1978,Lacey1993,Cole2000,Bell2003,Fontanot2009}. 

There have been many different density profiles of halos   developed to model the dark matter distribution in halos, and they are well reviewed by \citet{Keeton2001} and \citet{Cooray2002}. 
Among these models, the most commonly used model is the Navarro--Frenk--White (NFW) density profile \citep{NFW1996,Navarro1997}. 
There have been also some modified NFW profile models developed in past works, for example, the modified NFW profile in \citet{Maller2004} and the cuspy profile \citep{Moore1994,Kravtso1998,Moore1999}.  
Some other well known dark matter density profiles of haloes are the Hernquist profile \citep{Hernquist1990} and the singular isothermal sphere profile \citep{Sheth2001a}. In addition to model the dark matter distribution,  an  exponential profile \citep{Padmanabhan2017} has been used to model the distribution of  neutral atomic hydrogen (\HI)~gas  in a halo. 
Particularly, there have been also several works that have studied the dark matter halo shape of a single galaxy; for example, 
\citet{Brien2010} used
the pseudo-isothermal sphere profile  to model the dark matter halo of galaxy UGC 7321. However, in cosmology, we are only interested in the halo models that describe \emph{the distribution of galaxies} in groups and clusters, which need not match that of the total dark matter.

In cosmological large-scale structure analysis, mock catalogues of real galaxy surveys are required to estimate the errors of inferred cosmological parameters and identify possible systematics in the data and analysis methods. 
Mock surveys are often constructed with a halo occupation distribution model (HOD). In an HOD, a galaxy number density profile of haloes must be used to assign positions and velocities to mock galaxies inside the halo. 
In a significant number of previous works,  the galaxy number density profile of haloes has been assume to be the same as the dark matter density profile of haloes, or
the radial distribution of galaxies in a halo has been assumed to follow the total dark matter, modelled by the NFW profile \citep[e.g.][]{Guo2014,Guo2015,Howlett2015,Howlett2022,Qin2019b,Qin2021b,Alam2020,Avila2020,Paranjape2021}.

However, other recent research indicates that
galaxies tend to be found more in the outskirts of halos (nearer the virial radius) than an NFW profile, implying their distribution is not the same as the dark matter. {For example,  \citet{Avila2020} used the eBOSS survey to fit the HOD model, they found that  to optimize the HOD models to reproduce the galaxy clustering, the NFW profile has to be modified. 
Using simulations, \citet{Orsi2018} find that the emission line galaxies tend to be found in the outskirts of halos. Other examples can be found in \citet{Guo2014,Chen2017,Kraljic2018,Qin2022}.}
As a consequence, the measured galaxy two-point correlation function on small separation scales agrees poorly with the prediction from profile models of haloes developed in the past works, such as the NFW profile (\citealt{Qin2022}, hereafter \Q22).  
\citet{Chen2023} study the distribution of dark matte, rather than galaxies, in the outskirts of halos,  based on the dark matter simulation DarkEmu \citep{Nishimichi2019}, and this simulation does not account for any galaxy formation processes.

{   
In addition, the study of the `lensing-is-low' problem  indicates that HOD models optimized to reproduce
the clustering of galaxies over-predict their galaxy-galaxy lensing \citep{Leauthaud2017}. Using the IllustrisTNG \citep{Pillepich2018} simulation, 
\citet{Chaves2023} find that the origin of the lensing-is-low problem may due to the fact that
the  galaxies are
less concentrated than the prediction of NFW profile.
}

There has also been recent interest in the odd-multipole galaxy correlation function and power spectra   \citep{McDonald2009b,Bonvin2014,Gaztanaga2017,DiDio2019,Beutler2019}, such as the dipole and octopole of the correlation function/power spectrum. The higher multipole ($\ell>0$) statistics are generated by redshift-space distortion (RSD) effects, but the odd-multipoles are normally set to zero, assuming that RSD comes from only the quadratic Kaiser term. Non-zero odd-multipoles are generated by unequal gravitational potentials of different galaxies in the host halos, have been introduced as a method to constrain cosmological models. An incorrect galaxy number density profile of haloes will result in incorrect predictions of correlation function odd-multipole, and this will further bias the measurements of the cosmological parameters. Therefore, it is necessary to accurately model the galaxy number density profile of haloes.

In this paper, we use a simulated galaxy catalogue built with the \ds~semi-analytic model of galaxy formation \citep{Stevens2018} to explore the functional form of the galaxy number density profile of haloes. This simulation doesn't populate the halos with galaxies just using a particular density profile of haloes, but rather the distribution is a prediction as a result of numerous coupled galaxy evolution processes. We compare and validate this galaxy number density profile model of haloes against measurements from the group catalogue of Sloan Digital Sky Survey Data Release 10 \citep{York2000,Ahn2014}. Our galaxy number density profile of haloes will be used to generate mocks for the coming galaxy surveys, for example the Widefield ASKAP L-band Legacy All-sky Blind surveY (or WALLABY, \citealt{Koribalski2020}), the Dark Energy Spectroscopic Instrument (DESI, \citealt{DESI2016}) and the Square Kilometre Array (SKA, \citealt{2020PASA...37....7S}).   

 { In the past, there are less research on galaxy number density profile of haloes.}
 \citet{Budzynski2012}  only measured the projected (surface) galaxy density profiles \citep{Bartelmann1996} of halos (using SDSS DR7),   they did not develop any analytic models for the density profile and they only use the clusters with Luminous Red Galaxies at their centers for their measurements.

Our paper is structured as follows. In Section \ref{sec:datasim}, we introduce the {\sc Dark Sage} simulation and the SDSS DR10 group catalogue. In Section \ref{sec:hod}, we explore the functional form of the galaxy number density profile of haloes using \ds. In Section \ref{sec:Xigg}, we fit the galaxy two-point correlation function and constrain the HOD for \ds~galaxies to validate our galaxy number density profile of haloes. In Section \ref{sec:sdssFcs}, we validate  our galaxy number density profile of haloes using the SDSS DR10 group catalogue. 
We conclude in Section \ref{sec:conc}.

Hereafter, the capital letter $M \equiv \log_{10}(m/m_{\odot})$ denotes the logarithmic mass per unit solar mass $m_{\odot}$.

\section{Simulation and Data}\label{sec:datasim}

\subsection{The D{\small ARK} S{\small AGE} simulation} 

In this work, we use the {\sc Dark Sage} semi-analytic model of galaxy formation \citep{Stevens2016,Stevens2018} to explore the functional form of the galaxy number density profile of haloes. The simulated galaxy catalogue from \ds~used in this paper is the same as the simulation used in \Q22.

\ds~builds and evolves galaxies within halo merger trees constructed from an $N$-body simulation. 
The 2018 model used here was run on the widely
used Millennium simulation \citep{Springel2005}. This simulation 
assumes the $\Lambda$ cold dark matter model ($\Lambda$CDM) with cosmological parameters from
the Wilkinson Microwave Anisotropy Probe (WMAP)  First-Year results \citep{Spergel2003}, with matter density  $\Omega_{\rm m} = 0.25$, dark energy density $\Omega_{\Lambda}=0.75$, and Hubble parameter $h=0.73$ (where $H_{0} = 100$ $h$ km s$^{-1}$ Mpc$^{-1}$). The comoving box size of the simulation is $500\,h^{-1}$\,Mpc.

\ds~
accounts for  
gas accretion and cooling, 
star formation,
stellar feedback,
gravitational disc instabilities,
central black-hole growth, 
active galactic nuclei feedback,
environmental stripping of gas,
and  
galaxy mergers.  
Among other observational constraints, \ds~was calibrated to reproduce
the \HI-to-stellar mass ratio \citep{Brown2015} and  \HI~mass function \citep{Zwaan2005,Martin2010} with its eight free parameters~\citep{Stevens2018}. 
In \ds, all halos originally start with a galaxy at the centre, but the complicated merger history between halos of different masses as well as the secular and co-evolution of galaxies leads to some of those galaxies becoming satellites in larger halos with a radial dependence that is not easy to predict from first principles.
In this paper, we apply a halo mass cut $M=11.7$ to the \ds~catalogue to remove the halos that are below the simulation resolution.  {The halo mass range of \ds~ can cover the halo mass range of DESI $M\in[12,14.5]$ \citep{Wang2022,Yuan2023,Rocher2023}. In addition, as discussed  in Section 2 of \Q22, the halo mass range of \ds~ can also cover WALLABY.}

\subsection{The galaxy group catalogue from SDSS DR10}  
To measure the radial distribution of satellite galaxies in halos 
from real data for validation and comparison purposes, we use the galaxy group/cluster catalogue of \citet{Tempel2014}%
\footnote{Data can be downloaded from \url{https://cdsarc.cds.unistra.fr/viz-bin/cat/J/A+A/566/A1}}.
The group catalogue \citep{vizier:J/A+A/566/A1} is mainly based on the Sloan Digital Sky Survey (SDSS) Data Release 10 (DR10, \citealt{York2000, Ahn2014}).

The primary sample for the group catalogue was taken from the {\ttfamily SpecObj},  {\ttfamily bestobjid} and {\ttfamily fluxobjid} tables, which can be downloaded from the Catalogue Archive Server 
of the SDSS \citep{Tempel2014}.
Due to fiber collisions of the instrument, the SDSS galaxy sample is not complete \citep{Tempel2014}. Therefore, there are 1119 galaxies from the Two
Micron All Sky Survey (2MASS, \citealt{Jarrett2003}) Redshift Survey (2MRS \citealt{Huchra2012}), 
3494 galaxies from the Two-degree Field Galaxy Redshift Survey (2dFGRS, \citealt{Colless2001b,Colless2003}) as well as 280 galaxies from the Third Reference Catalogue of Bright Galaxies (RC3, \citealt{Corwin1994}) that have been added to complement the SDSS sample. 
The final sample totals 588\,193 galaxies, upon which the group catalogue was constructed. 

A friends-of-friends (FoF) algorithm was implemented by \citet{Tempel2014} to identify the galaxy groups.  
{In this paper, we use a volume-limited catalogue of  \citet{Tempel2014}. As mentioned in \citet{Tempel2014}, since in the volume-limited catalogue, the number density of galaxies is a constant by definition, therefore should be used to make comparison with simulations.}  

In the volume-limited catalogues,
the absolute magnitude of galaxies was transformed from their apparent magnitude using the group mean redshift \citep{Tempel2014}. A 
$k$-correction and evolution-correction have also been applied to the data. To suppress the redshift-space distortion (RSD, or the Finger-of-God effect, \citealt{Kaiser1987}), the group mean distances were assigned to their member galaxies.
The 3D Cartesian comoving positions of the galaxies have been accordingly calculated \citep{Tempel2014}. 
These positions will be used to measure the halo density profile for galaxies in our paper. 
The line-of-sight peculiar velocities of galaxies were determined from the group mean distances and the observed redshift of the galaxies \citep{Tempel2014}.  

\begin{figure} 
\centering
 \includegraphics[width=\columnwidth]{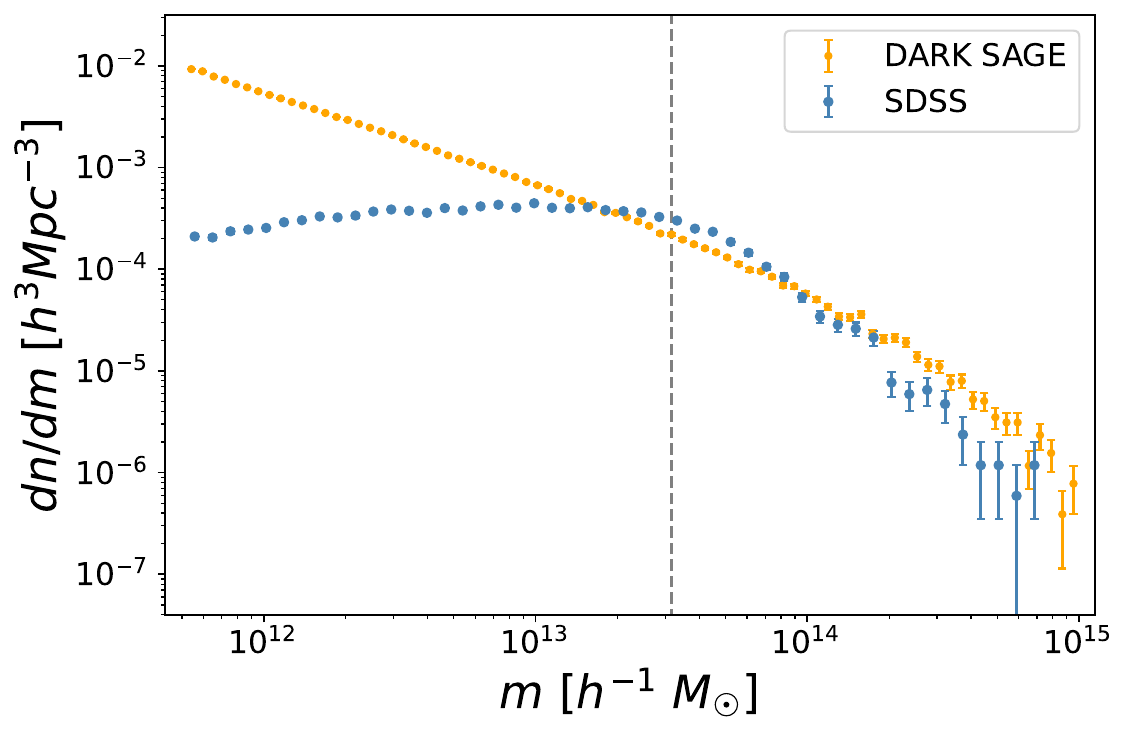}
 \caption{The halo mass function of SDSS (blue dots) and \ds~(yellow dots). Here we only intend to show the turning point of the halo mass function of SDSS.} 
 \label{halomf}
\end{figure}

Assuming that galaxy groups inhabit common dark matter halos, 
the total mass of  a group (or halo) is calculated using 
\be 
m=2.325\times10^{12}\frac{R_g}{\rm Mpc}\left(\frac{\sigma_v}{\rm 100 ~km~s^{-1}}\right)^2m_{\sun} \,,
\ee 
which is derived from the virial theorem \citep{Tempel2014}. In the above equation, 
$\sigma_v$ is the velocity dispersion (reflecting the average kinetic energy of the group members) which was estimated using the line-of-sight peculiar velocities of galaxies and the mean group velocity as well as the mean group redshift \citep{Tempel2014}. 
$R_g$ is the gravitational radius (reflecting the average potential energy of the group members), which was calculated by assuming the {\it total} matter (rather than galaxies) in a halo is modelled by the NFW profile \citep{Bartelmann1996,Lokas2001,Tempel2014}. 
In the catalogue, group masses range from $10^9\,m_{\sun}$ to $10^{15}\,m_{\sun}$. 
We plot the halo mass function of SDSS data in Fig.\ref{halomf}, as shown by the yellow dots, the halo mass function of SDSS drops down for $M<=13.5$ since 
the catalogue is not complete due to 
the fibre collision effects and luminosity limits in the observation. Therefore, in this paper, 
we only use the halos with $m>10^{13.5}\,m_{\sun}$ to measure the galaxy number density profile of haloes.

\citet{Tempel2014} created seven volume-limited catalogues with different absolute ($r$-band) magnitude cuts. 
In this paper, we use the catalogue with magnitude cut $-20.0$ since the redshift cut of this catalogue is 0.11, which is comparable to the redshift limit of  WALLABY. 
There are 163\,094 galaxies and 24\,258 groups in the catalogue. The LL of the FoF algorithm was set to be 0.515\,$h^{-1}$ Mpc \citep{Tempel2014}.  


The smallest groups in our data sample of choice  contain only two galaxies. Around  52.1\% of the galaxies are `isolated' galaxies, meaning they are the sole occupant of their halo with no satellite galaxies. Since we want to measure the radial distribution of satellite galaxies in halos, 
the isolated galaxies are excluded in our research. In the catalogue, the galaxies with luminosity {\ttfamily rank}~%
equal to one (the most luminous galaxies) are identified as the central galaxies \citep{Tempel2014}. 

\section{The galaxy number density profile of haloes} \label{sec:hod}

\subsection{The galaxy number density profile model of haloes}

In the course of \Q22, we found that radial distribution
of satellite galaxies within \ds~ was considerably wider than predicted by the NFW profile. As a result of that finding, assuming the halo is spherically symmetric\footnote{{For an HOD model, we’re interested in capturing the average behaviour of galaxy positions, if stacked the satellite positions of a bunch of haloes, we would end up with a spherically symmetric distribution. The assumption of spherical symmetry
is good enough for HOD.}}, in this work we instead adopt the following function to model the satellite galaxy number density profile of haloes:
\be \label{newprof}
\rho_g(r)  \propto  r^2 \exp\left[ - \beta (cr)^\alpha \right]\,,
\ee
where $c\equiv R_{\mathrm{vir}} / R_s$ is the halo concentration. $R_s$ is the  break  radius  between  the  outer and inner density  profile of the halo. $R_{\rm vir}$ is the virial radius of a halo, it is computed from the halo virial mass $m$ using 
\be \label{Rvirs}
R_{\mathrm{vir}}=\left( \frac{3 m}{4\uppi\, \Delta\, \rho_c}\right)^{1/3} ~,
\ee 
where $\Delta=200$ is the overdensity threshold of halo identification. $\rho_c=3H_0^2/(8\uppi G)$ is the  critical density of the  universe. 
$G$ is Newton's gravitational constant.   
The parameters $\alpha$ and $\beta$ are functions of halo (virial) mass. 
Assuming the central galaxy sits at the center of potential of the halo,
the radial distribution (probability density function) of satellite galaxies in a parent halo is then given by
\be\label{fcseq}
f_{cs}(x)\propto \rho_g(r)\, r^2
\ee
\citep{Zheng2004,Tinker2005,Zheng2007}.
$f_{cs}(x)$ should be normalized to 1 when integrating from $x=0$ to $x=1$, where $x\equiv r/(10R_{\rm vir})$. The number 10 is chosen to ensure the interval $[0, 1]$ is large enough to cover the whole shape of the curves (see section 5.1 of \Q22 for more discussion).

\begin{figure} 
\centering
 \includegraphics[width=\columnwidth]{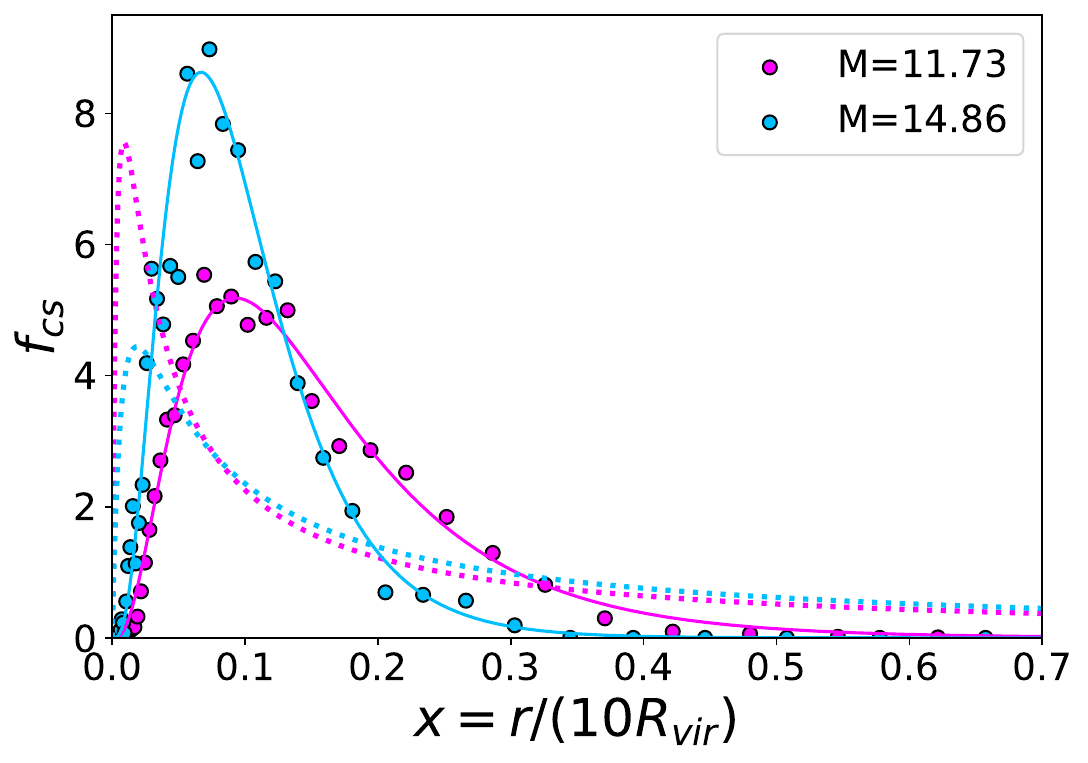}
 \caption{The {probability density function (the radial distribution) of satellite galaxies} $f_{cs}$ for two example halo mass bins.  The pink and blue filled circles are the measurements from {\sc Dark Sage} for halo mass bins centered on $M=11.73$, $14.86$ respectively. 
 The solid curves are the models fit to the measurements.  
 The dashed curves compare an NFW profile for these masses. All of them are normalized to integrate to one in the interval  $x\in[0,1]$. } 
 \label{fitfcs}
\end{figure}

\begin{figure} 
\centering
 \includegraphics[width=\columnwidth]{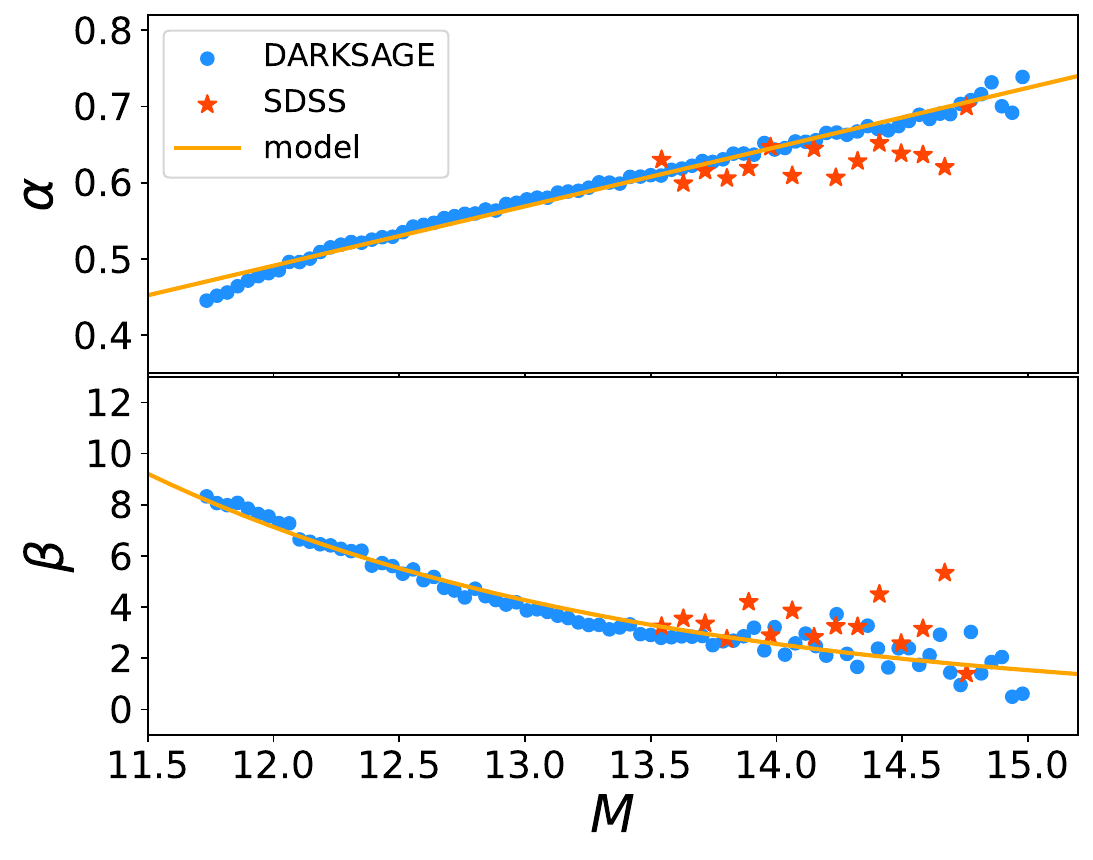}
 \caption{Top panel: blue dots show the $\alpha$  fitted by comparing
the halo density profile model for galaxies Eq.\ref{newprof} to the measurements of \ds~in each halo mass bin. 
The yellow curve is the model Eq.~\ref{Am} fit to the blue dots. Similarly, in the bottom panel, the blue dots show the fitted $\beta$ of the measurements of \ds~satellites in each halo mass bin, and the yellow curve is the Eq.~\ref{Bm} fit to the blue dots. In both cases, the red pentagrams are the results from SDSS data (see Section~\ref{sec:sdssFcs}).} 
 \label{fitAB}
\end{figure}

To explore the functional forms of $\alpha$ and $\beta$, 
we fit our model Eq.~\ref{newprof} to the measurements from \ds~ in 80 halo mass bins  
in the interval $M\in[11.7,15]$ {(the mass bin width is around 0.04)}.
We plot the measured $f_{cs}$ against $x$ in  Fig.~\ref{fitfcs} for two example halo mass bins at each extreme of the range. The solid curves are the models fit to the measurements.  All these $f_{cs}$ are normalized
to integrate to one in the interval $x\in[0,1]$. 
The fit results of $\alpha$ and $\beta$ against the halo mass are shown in the blue dots of Fig.~\ref{fitAB}.
We find a linear model fits the relation between $\alpha$ and halo mass $M$ very well;
\be \label{Am}
\alpha= 0.0778 M - 0.4419\,.
\ee 
The fit result is shown in the yellow line in the top panel of Fig.~\ref{fitAB}.
We use an exponential function to model the relation between $\beta$ and $M$, whose best fit is given by
\be \label{Bm}
\beta= 3331.4392\exp\left[-0.5123M\right]\,,
\ee 
This fit result is illustrated by the yellow curve in the bottom panel of Fig.~\ref{fitAB}. 
In all the above fitting procedures, we fit the measurements to the model curves (lines) by simply minimizing the least squares difference. The combination of Eq.~\ref{newprof}, \ref{Am} and \ref{Bm} comprise the density profile model we will adopt in this paper.

In Fig.\ref{fitfcs}, 
the dashed curves
depict the NFW profile for these masses assuming an one-to-one mass-concentration relation. The measured $f_{cs}$ of the galaxies from {\sc Dark Sage} clearly does not agree with an NFW profile. 
\ds~ gives satellite galaxies that are found much closer to or beyond the Virial radius than would be predicted by the more centrally concentrated NFW profile. 

\begin{figure*} 
\centering
 \includegraphics[width=89mm]{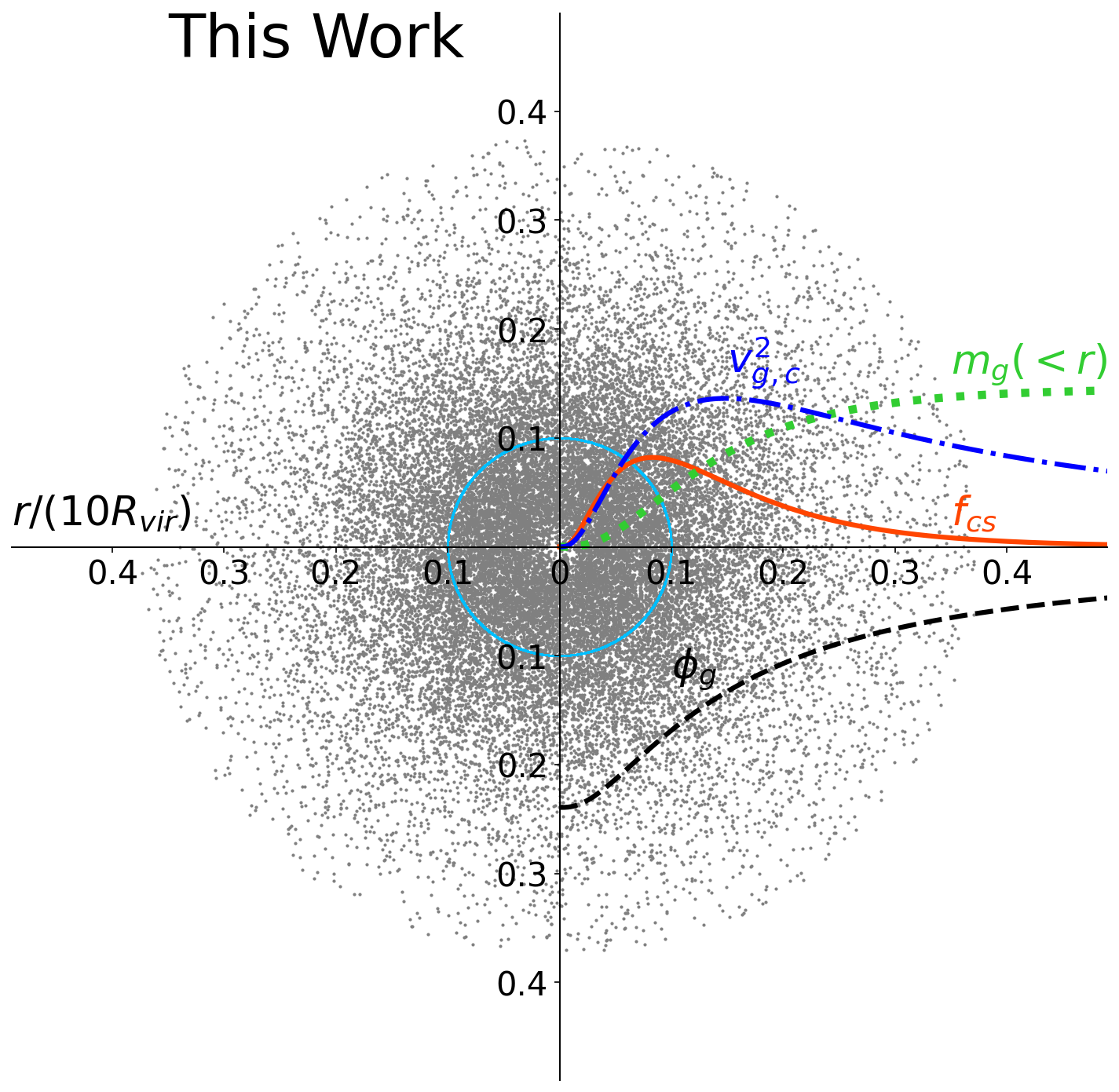}
  \includegraphics[width=89mm]{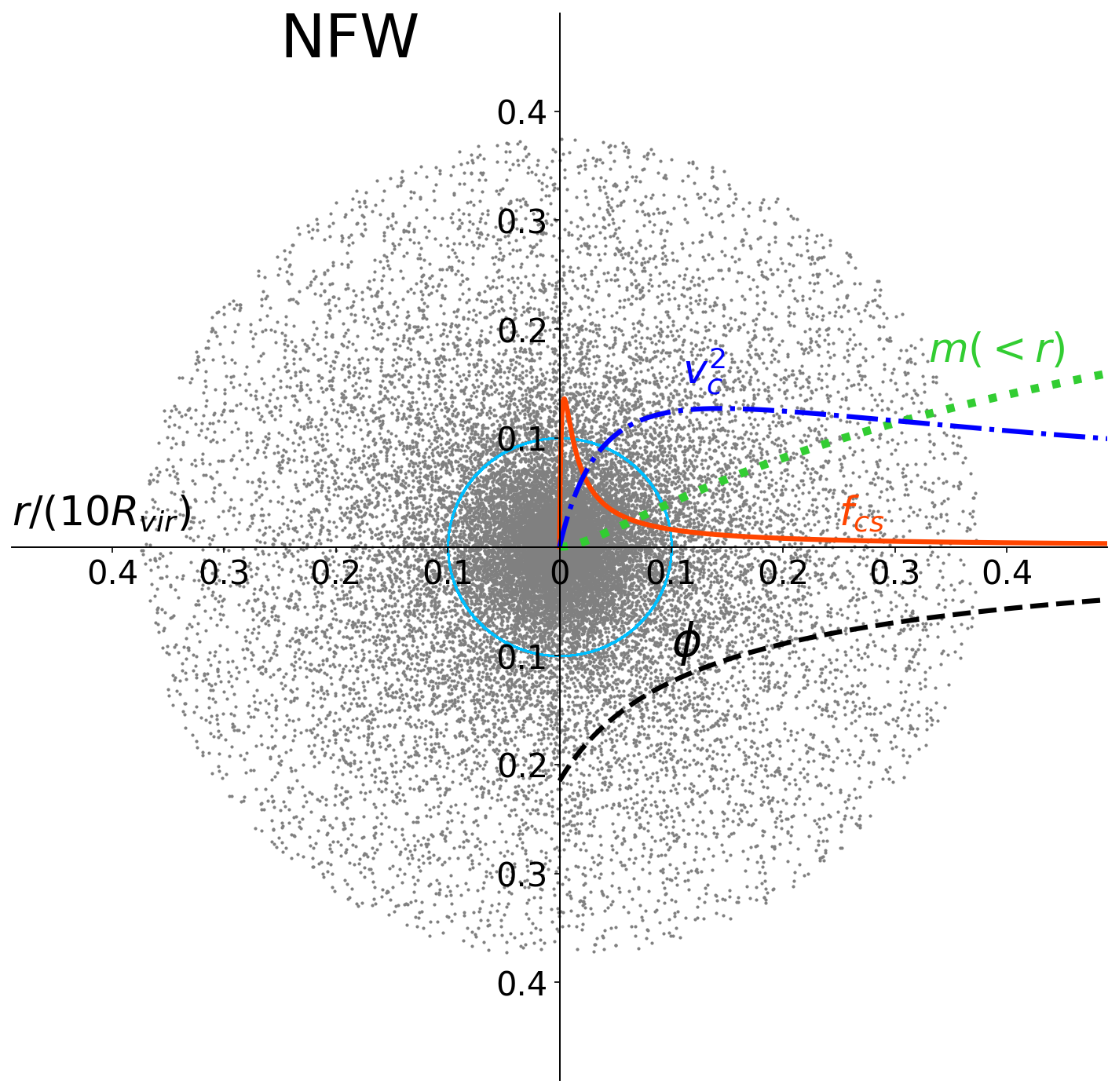}
 \caption{
 Comparing the   profile model of Eq.~\ref{newprof} (left) to the NFW profile (right). The gray dots are random points generated based on these two density profile models for halo mass $M=12$. The light-blue circles indicate the  virial
radius of the halo mass $M=12$. The blue dashed-dotted curves illustrate the circular velocity $v^2_c$.
The green dotted curves show the mass  interior
to $r$, i.e. $M(<r)$.
The yellow dashed curves display the gravitational potential  $\phi$. The red solid curves showcase the $f_{cs}$.  The amplitudes of the curves are re-scaled in order to be visible in the same figure. The coordinates axis indicate the radial distances from the halo center, and are re-scaled to $r/(10R_{\rm vir})$. 
} 
 \label{dynas}
\end{figure*}

As a more intuitive visual representation, we show in the left-hand panel of Fig.~\ref{dynas} gray dots that are randomly generated assuming our new galaxy number density profile model of haloes for $M=12$. 
The red curve illustrates the corresponding $f_{cs}$ of this  halo mass. The light-blue circle indicates the position of the virial radius $R_{\rm vir}$ of the halo. 
Under our new model, satellite galaxies tend to be situated more in the outskirts of a halo.
For comparison, the gray dots in the right-hand panel are generated assuming the NFW profile for the same halo mass (see Appendix \ref{sec:nfw} for relevant equations). 
Under the NFW profile, satellite galaxies are majorly concentrated within the virial radius and show no suppression at the halo center.

 There is also a physical motivation for a satellite PDF that peaks closer to $R_{\rm vir}$ than $R=0$. Galaxy orbits with a smaller radius relative to the centre of mass are not stable, as the galaxies are subject to a stronger tidal field and greater dynamical friction, meaning they will merge with the central galaxy and/or become stripped/disrupted over relatively short time-scales, and therefore cease to exist. However, though we can model these process using simulations, there are still challenges involved in identifying subhalos near halo centres in these simulations. The functional form we have selected in Eq.~\ref{newprof} is a phenomenological one based on \ds, but we will later test/validate it against radial profiles of galaxy groups from SDSS data.
For now, we have tested that this is not a specific feature of the Millennium simulation.

\subsection{The convolution and Fourier transform for the galaxy number density profile of haloes}

The probability density function (PDF) of satellite
galaxy pairs in a parent halo is calculated from 
the
convolution of   a profile with another of exactly the same
shape, given by \citep{Sheth2001a,Zheng2007}
\be \label{convs}
f_{ss}(r)\, \diff r \propto r^2\, \diff r \int_{0}^{+\infty} \left[  \tau_1^2 \rho_g(\tau_1)\int_{-1}^{1}\rho_g(\tau_2)\,\diff\mu \right] \diff\tau_1\,,
\ee 
where
\be
\tau_2=\sqrt{\tau_1^2+r^2-2\tau_1 \mu}
\ee 
and ${\bf r}=\boldsymbol{ \tau}_2-\boldsymbol{ \tau}_1$ is the pair separation between position vectors $\boldsymbol{  \tau}_1$ and $\boldsymbol{ \tau}_2$.  $\mu=-\cos\theta$, where $\theta$ is the angular between 
${\bf r}$ and $\boldsymbol{ \tau}_1$.  $f_{ss}(x)$ should be normalized to 1 when integrating from $x=0$ to $x=1$.

Plugging Eq.~\ref{newprof} into Eq.~\ref{convs}, one can obtain
\begin{multline}
\label{convnew}
f_{ss}(r) \propto \frac{\beta^{-4r/\alpha}}{\alpha c^4}\int_{0}^{+\infty}\tau_1^3\exp[-\beta(c\tau_1)^{\alpha}]~ \times \\
\left[\Gamma\left(\frac{4}{a},~\beta c^{\alpha}|r-\tau_1|^{\alpha}\right)-\Gamma\left(\frac{4}{a},~\beta c^{\alpha}|r+\tau_1|^{\alpha}\right) \right] \diff\tau_1\,,
\end{multline}
where the regularized upper incomplete $\Gamma$ function is defined as:%
%
\be
\Gamma(A,x)\equiv\frac{1}{\Gamma(A)}\int_x^{+\infty}t^{A-1}\exp(-t) \,\diff t
\ee
and where the complete $\Gamma$ function $\Gamma(A)\equiv \Gamma(A,0)$. 
Due to the $\Gamma$ functions inside the integral of Eq.~\ref{convnew}, we can not solve the integration analytically. However, it can be solved using numerical methods easily; we first calculate a spline function for the term inside the integral, then calculate the integral numerically. 

\begin{figure} 
\centering
 \includegraphics[width=\columnwidth]{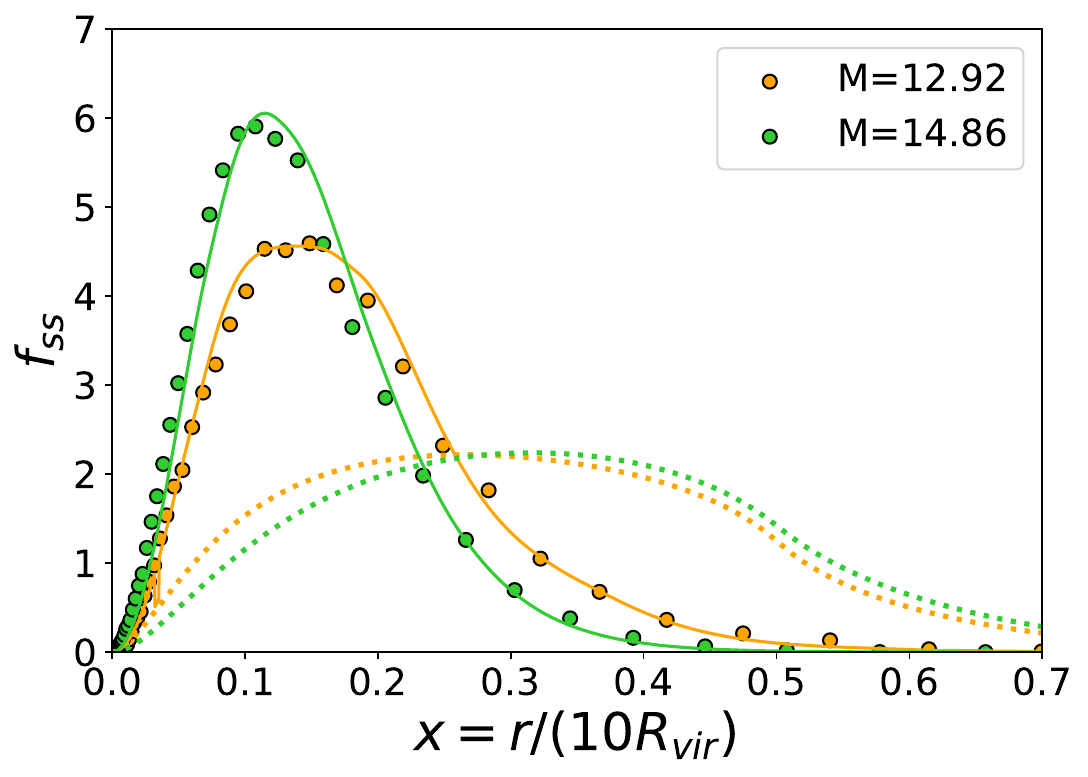}
 \caption{The {probability density function (the radial distribution) of satellite
galaxy pairs} $f_{ss}$ for two example halo mass bins.  The yellow and green filled circles  are the measurements from {\sc Dark Sage} for halo mass bins centered on $M=12.68$, $14.86$ respectively. 
 The solid curves are the model calculated from Eq.~\ref{convnew}, using the mass-dependent parameter values for $\alpha$ (Eq.~\ref{Am}) and $\beta$ (Eq.~\ref{Bm}). 
 The dashed curves showcase the NFW profile for these masses. All of them are normalized to integrate to one in the interval  $x\in[0,1]$. } 
 \label{fitfss}
\end{figure}

We plot the measured $f_{ss}$  {(i.e. the PDF of satellite-satellite pairs  measured in halo mass bins)}  against $x$ in Fig.~\ref{fitfss}, as shown in the circles for two example halo mass bins. 
The solid curves represent our new model.
While the free parameters of our model were fitted to \ds, the curves in Fig.~\ref{fitfss} are \emph{not} direct fits to the points in Fig.~\ref{fitfss}.
This figure therefore validates the general applicability of our model with its fitted parameters. 
The dashed curves depict the NFW profile for these mass bins, which, again, clearly do not match the simulated data. All these $f_{ss}$ are normalized to integrate to one in the interval $x\in[0,1]$. 

\begin{figure} 
\centering
 \includegraphics[width=\columnwidth]{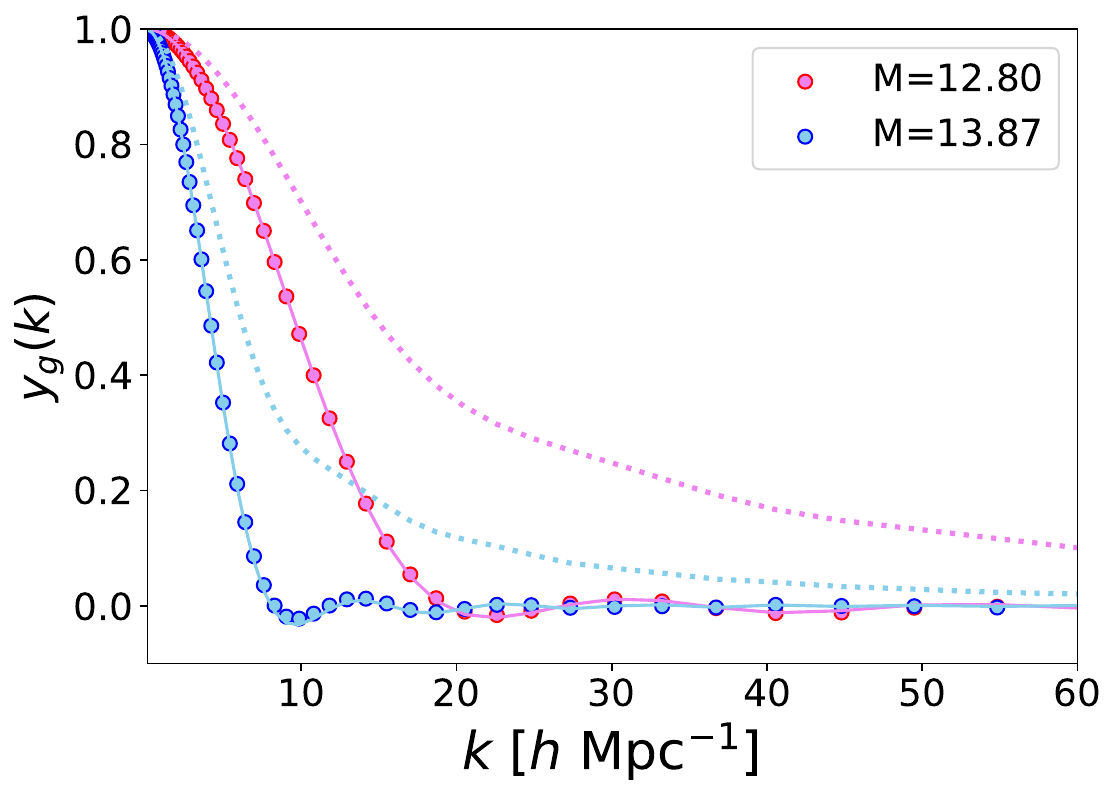}
 \caption{The {Fourier transformation for the galaxy number density profile of haloes}, in two example halo mass bins.  The pink and blue filled circles  are the measurements from {\sc Dark Sage} for halo mass bins centered on $M=12.80$, $13.87$ respectively. 
 The solid curves are the model calculated from Eqs~\ref{ygeq} \& \ref{newprof},  using the mass-dependent parameter values for $\alpha$ (Eq.~\ref{Am}) and $\beta$ (Eq.~\ref{Bm}). 
 The dashed curves showcase the NFW profile for these masses. } 
 \label{fityg}
\end{figure}

The Fourier transformation for the galaxy number density profile of haloes is defined as \citep{Cooray2002}
\be \label{ygeq}
y_g(k)\equiv\int_0^{R_{\mathrm{vir}}} 4\uppi R^2 \frac{\sin(kR)}{kR}\frac{\rho_g(R)}{m}\, \diff R \,.
\ee   
Plugging Eq.~\ref{newprof} into the above, 
one can be calculate $y_g(k)$ numerically. 
We plot the measured $y_g$ in  Fig.~\ref{fityg}, as shown in the circles for two example halo mass bins.  
The solid curves compare the models directly calculated from Eqs~\ref{ygeq}, \ref{newprof}, \ref{Am} and \ref{Bm}. They  agree with the measurements.  
This figure therefore validates the general applicability of our model with its fitted parameters. 
For reference, the dashed curves depict the NFW profile for these mass bins. 
All these $y_g$ are normalized to 1 at $k=0$.

\subsection{The dynamic properties} 

The galaxy number density profile of haloes of Eq.~\ref{newprof} is originally developed to model the distribution of galaxies. The dark matter in a halo is still assumed to be modeled by the NFW profile. The dynamic properties inferred from Eq.~\ref{newprof} can be treated as a perturbation term added upon the dynamic properties of the NFW profile. In this section, we will explore the dynamic properties corresponding to Eq.~\ref{newprof}. For comparison, we also describe the dynamic properties of the NFW profile in Appendix \ref{sec:nfw}.

If $m_g$ denotes the mass of galaxies in a halo, we can normalize Eq.~\ref{newprof}   to   mass $m_g$ in the interval $r\in[0,~+\infty]$. The normalized galaxy number density profile of haloes is given by  \be \label{newprof2}
\rho_g(r) = \frac{\alpha \beta^{\frac{5}{\alpha}}c^5m_g}{4\uppi \Gamma[\frac{5}{\alpha}]}  r^2 \exp\left[ - \beta (cr)^\alpha \right]  \,.
\ee

The circular velocity at a radial distance
$r$ 
from the halo center is
defined as
\citep{Sheth2001a}
\be \label{vcnew}
v^2_c\equiv\frac{G m_g(<r)}{r}\,,
\ee
where the mass interior to $r$ is given by
\be 
m_g(<r)=4\uppi\int_0^{r}\tau^2\rho_g(\tau)d\tau\,.
\ee
Plugging Eq.~\ref{newprof2} into the above equation yields 
\be  \label{Msmallr}
m_g(<r)=\frac{m_g}{\Gamma[\frac{5}{a}] } \left[\Gamma\left(\frac{5}{a}\right)-\Gamma\left(\frac{5}{a},~\beta c^{\alpha}r^{\alpha}\right) \right]\,.
\ee  
Plugging the above into Eq.~\ref{vcnew}, one can obtain
\be  \label{criclV}
v^2_{c,g}=\frac{ G m_g}{ r \Gamma[\frac{5}{a}]} \left[\Gamma\left(\frac{5}{a}\right)-\Gamma\left(\frac{5}{a},~\beta c^{\alpha}r^{\alpha}\right) \right]\,,
\ee
which is the circular velocity for  Eq.~\ref{newprof2}.  In the left-hand  panel of Fig.~\ref{dynas}, the green-dotted curve displays the mass interior to $r$, the blue dash-dotted curve showcases the circular velocity. 
For comparison,  the green-dotted curve in the right-hand panel shows the mass interior to $r$ for  NFW profile. The blue dash-dotted curve showcases the circular velocity for  NFW profile.

The gravitational potential energy at
  a radial distance
$r$ 
from the halo center is
  given by
\citep{Sheth2001a}
\be
\phi_g(r)=\frac{-4\uppi G}{r}\int_0^{r}\tau^2\rho_g(\tau)d\tau-4\uppi G\int_r^{+\infty}\tau\rho_g(\tau)d\tau
\ee
Plugging Eq.~\ref{Msmallr} and Eq.~\ref{newprof2} into the above equation, one can obtain
\be\label{gwphi}
\begin{split}
\phi_g(r)=&-\frac{  G 
m_g}{r \Gamma[\frac{5}{a}]} \bigg [
\Gamma\left(\frac{5}{\alpha}\right)+
\beta^{\frac{1}{\alpha}}cr\Gamma\left(\frac{4}{\alpha},~\beta c^{\alpha}r^{\alpha}\right)   \\
&-\Gamma\left(\frac{5}{\alpha},~\beta c^{\alpha}r^{\alpha}\right)
 \bigg ]
\end{split}
\ee
which is the potential energy for  Eq.~\ref{newprof2}. An example is illustrated by the black dashed curve of the left-hand-side panel of Fig.~\ref{dynas}. 
For comparison,  the black dashed curve in the right-hand-side panel shows the gravitational potential for  NFW profile.

\section{Validating the galaxy number density profile of haloes  using the {\sc Dark Sage} simulation} \label{sec:Xigg}

In this section we will explore whether or not our new galaxy number density profile model of haloes can correctly reproduce the galaxy two-point correlation function of \ds~galaxies. 

\subsection{The model of HOD and correlation function}

In the HOD-based model, the galaxy two-point correlation function is computed as the
combination of the `one-halo' and `two-halo' terms
\citep{Berlind2002,Yang2003,Zheng2004,Tinker2005,Guo2015,Zheng2016,Qin2022}
\be \label{realxigg}
\xi_{gg}^{\rm mod}(r)=[1+\xi^{1h}_{gg}(r)]+\xi^{2h}_{gg}(r)~,
\ee 
where $r$ is the pair separation. The analytic expression of $\xi_{gg}^{\rm mod}$ in terms of an HOD model (or HOD parameters) and density profile model of haloes is fully presented in Section 4 of \Q22 (or see Appendix B of \citealt{Tinker2005}).  {The expressions are lengthy, we therefore only briefly introduce them in the following texts rather than displaying the full expressions.} 

The `one-halo' term $\xi^{1h}_{gg}$ is the contribution of intrahalo galaxy pairs, meaning it is sensitive to the halo density profile.   
{
The one-halo term is modeled as  
\be  
\begin{split}
 1+\xi^{1h}_{gg}(r)=&\frac{1}{  2\uppi\, r^2\, \bar{n}_g^2 }\int_0^{\infty} \frac{1}{10 R_{\mathrm{vir}}}
  \bigg [
 \left\langle N_{\mathrm{cen}}\, N_{\mathrm{sat}}\right\rangle\, f_{cs}(x) \\
 +&
 \frac{\left\langle N_{\mathrm{sat}}(N_{\mathrm{sat}}-1)\right\rangle }{2} f_{ss}(x)
 \bigg ] f_{h}(m)\, dm ~,
 \end{split}
\ee 
where $\bar{n}_g$ is the average number density of galaxies, $f_h(m)$ is the model of the halo mass function.}

The `two-halo' term $\xi^{2h}_{gg}$ is the contribution of interhalo galaxy pairs and is therefore \emph{not} sensitive to the halo density profile.{ The two-halo term is modeled as
\be 
\xi^{2h}_{gg}(r)= \left[1+\xi'_{2h}(r)\right]\left( \frac{\bar{n}'_g}{\bar{n}_g} \right)^2 -1 ~,
\ee 
where
\be  
\xi'_{2h}(r)=\frac{1}{2\uppi^2}\int_0^{\infty} b_{g}^{2}(k,r)\, P_m(k)\, k^2\, \frac{\sin(kr)}{kr}\, dk ~,
\ee 
and where 
$\bar{n}'_g$ is the average number density of galaxies that reside in halos with mass smaller than the so-called `halo exclusion' limit. 
$b_{g}$ is galaxy biasing parameter. $P_m(k)$ is the nonlinear matter power spectrum. }

The HOD of central galaxies of {\sc Dark Sage} is given by $\langle N_{\mathrm{cen}} \rangle =1$ since    {\sc Dark Sage} requires that each parent halo hosts a central galaxy. We choose  the following formula to model the HOD of satellite galaxies of {\sc Dark Sage} 
\citep{Zheng2007b,Howlett2015,Guo2015,Zheng2016}
\be \label{satHODDS}
\langle N_{\mathrm{sat}}  \rangle = \langle N_{\mathrm{cen}}  \rangle \left( \frac{m-10^{M_{\mathrm{cut}}}}{10^{M_1}} \right)^{\beta} ~,
\ee 
where the HOD parameters $M_1$, $\beta$ and $M_{\mathrm{cut}}$ will be fit by comparing the model correlation function $\xi_{gg}^{\rm mod}$  to the measurement $\xi_{gg}^{\rm mea}$.

In addition to HOD model, the galaxy number density profile of haloes is 
also the ingredient of $\xi_{gg}^{\rm mod}$. The 
$f_{cs}$ and $f_{ss}$ are used to model the one-halo term, while $y_g$ is used to compute the two-halo term. These are all computed using the expressions in Section~\ref{sec:hod}. Although some of them need to be computed numerically, they only need to be computed once and tabulated as a function of halo mass, and therefore do not add much computational time to calculating the model correlation function when fitting the HOD.


\subsection{Fitting results} 
 
\begin{figure} 
\centering
 \includegraphics[width=81mm]{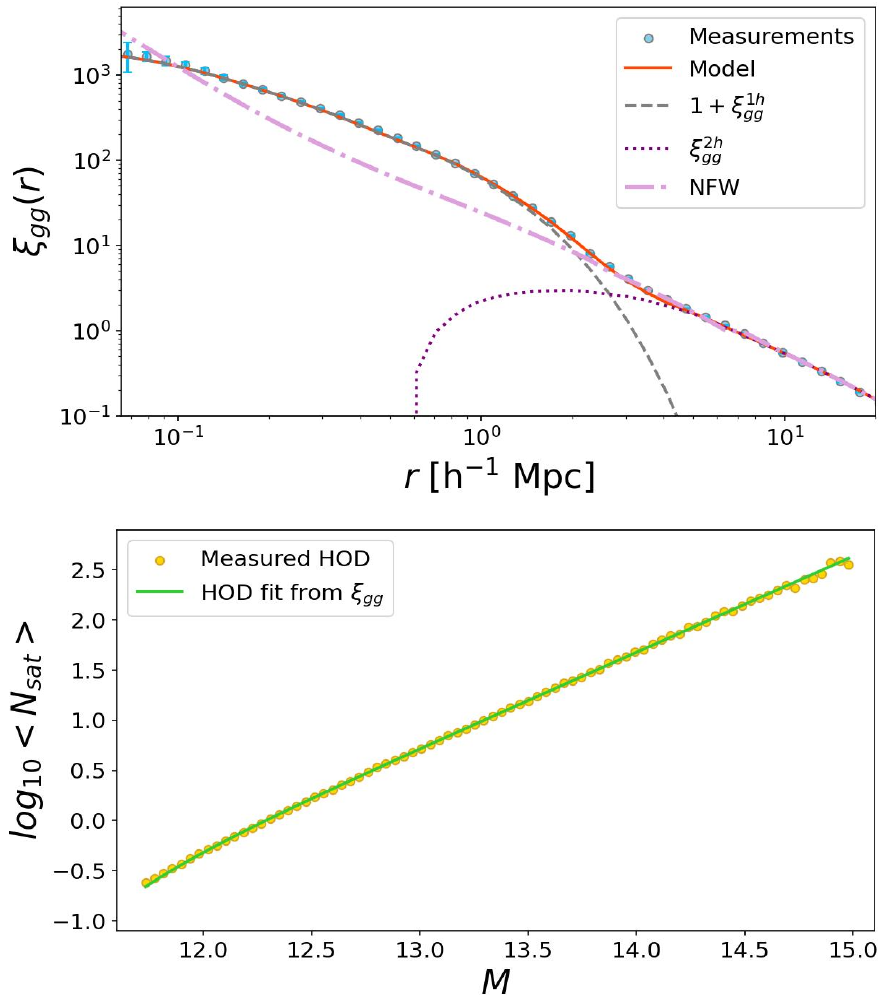}
 \caption{The top panel
shows the correlation functions. The blue-filled circles (for which the error bars are small enough to be unnoticeable except on very small scales) are the measured correlation function using the real-space Cartesian coordinates of the \ds~ galaxies. The red solid curve is the model correlation function fit to the measurements using our new halo density profile for galaxies. The pink dash-dotted curve is the model correlation function calculated using the NFW profile and the yellow dots of the bottom panel. In the bottom panel, the green line is the HOD \textit{inferred} from the fitted correlation function using our new profile model, the yellow dots are the measured HOD ({errorbars are tiny}). So our fitting method combined with the new profile model does an excellent job of recovering the true HOD from fitting only the correlation function. The fit results of the HOD parameters are displayed in Fig.\ref{xigghod222} of Appendix \ref{sec:hodp}. 
} 
 \label{xigghod}
\end{figure}

The top-left panel of Fig.~\ref{xigghod} shows several correlation functions. 
The blue-filled circles (which have negligible errors for most points) are the measured correlation function using the real-space Cartesian coordinates of the galaxies in \ds. The red solid curve is the model correlation function fit to the measurements. The pink dash-dotted curve is the model correlation function calculated using the NFW profile (and the measured HOD of \ds). We also show the measured HOD and best fitting model using our new galaxy number density profile of haloes in the bottom-left panel. 

Comparing the pink dash-dotted curve to the blue filled circles and red solid curve, we find that at larger scales $r>3\,h^{-1}$ Mpc, the model correlation function calculated using the NFW profile agrees with the measurements and the model predicted by Eq.~\ref{newprof}, since this scale is dominated by the two-halo term, which is not sensitive to the galaxy number density profile of haloes. 
However, at smaller scales, i.e. $r<3\,h^{-1}$ Mpc,
a galaxy number density profile of haloes different to NFW, such as the one developed in this work, is needed to correctly recover the correlation function.
%
Particularly, at scales of $r<0.1\,h^{-1}$ Mpc, the model correlation function calculated from the NFW profile is too high compared to the measurements, as there are fewer galaxies situated in the halo center region (see the left-hand panel of Fig.~\ref{dynas} for an example illustration).
At scales of $r\in[0.1,~3]h^{-1}$ Mpc, the model correlation function calculated from the NFW profile is too low compared to the measurements, as satellite galaxies tend to be situated farther in the outskirts of the halo than the NFW profile predicts.
In the bottom-left panel of Fig.~\ref{xigghod}, the green line is the HOD inferred from the fitted correlation function (red solid curve), which is in excellent agreement with the measured HOD. The fit results of the HOD
parameters are displayed in the right-hand panels.  

To summarise, in this section, we have proven that the galaxy number density profile of haloes Eq.~\ref{newprof}, \ref{Am} and \ref{Bm} can accurately reproduce the HOD and correlation function of \ds~galaxies.

\section{Validating the galaxy number density profile of haloes using SDSS galaxies}\label{sec:sdssFcs}

\subsection{Comparing the   galaxy number density profile model of haloes to the measurements from SDSS galaxies}\label{sec:5.1}

In Fig.~\ref{fitAB}, the red pentagrams show the $\alpha$ and $\beta$ values against halo mass $M$, obtain by comparing Eq.~\ref{newprof} to the measurements of real SDSS galaxies in halo mass bins. {The halo mass bin width is around 0.06.}
The $\alpha$ values estimated from the SDSS galaxies are systematically lower than the values estimated from \ds ~(blue dots), while the $\beta$  values estimated from the SDSS galaxies are systematically higher than \ds ~(blue dots).
However, the functional forms of Eq.~\ref{Am} and \ref{Bm} can still be used to model the red pentagrams (with different coefficients). 
Due to limited measurements in lower halo mass bins, the SDSS points become too noisy to infer anything concrete about their distributions.
As such, we do not fit the them using Eq.~\ref{Am} and \ref{Bm}.  
In future work, we aim to further test this using larger galaxies surveys such as WALLABY and DESI.

\begin{figure} 
\centering
 \includegraphics[width=\columnwidth]{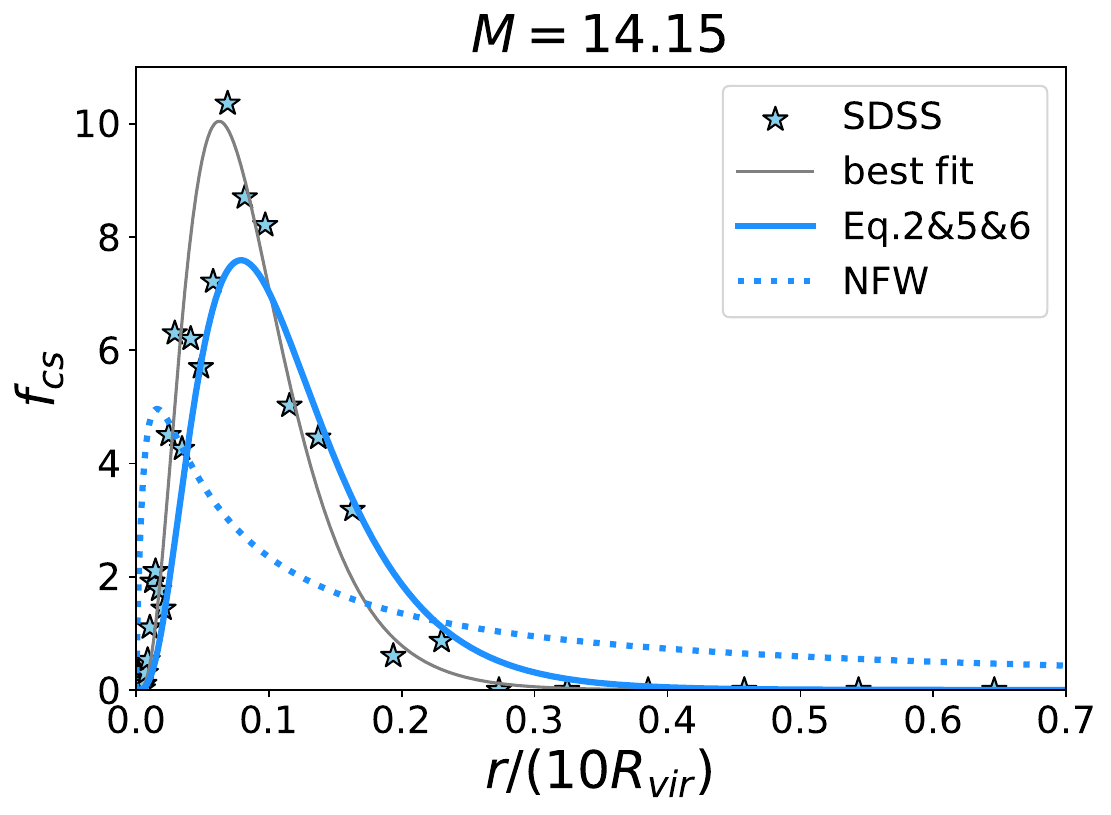}
  \includegraphics[width=\columnwidth]{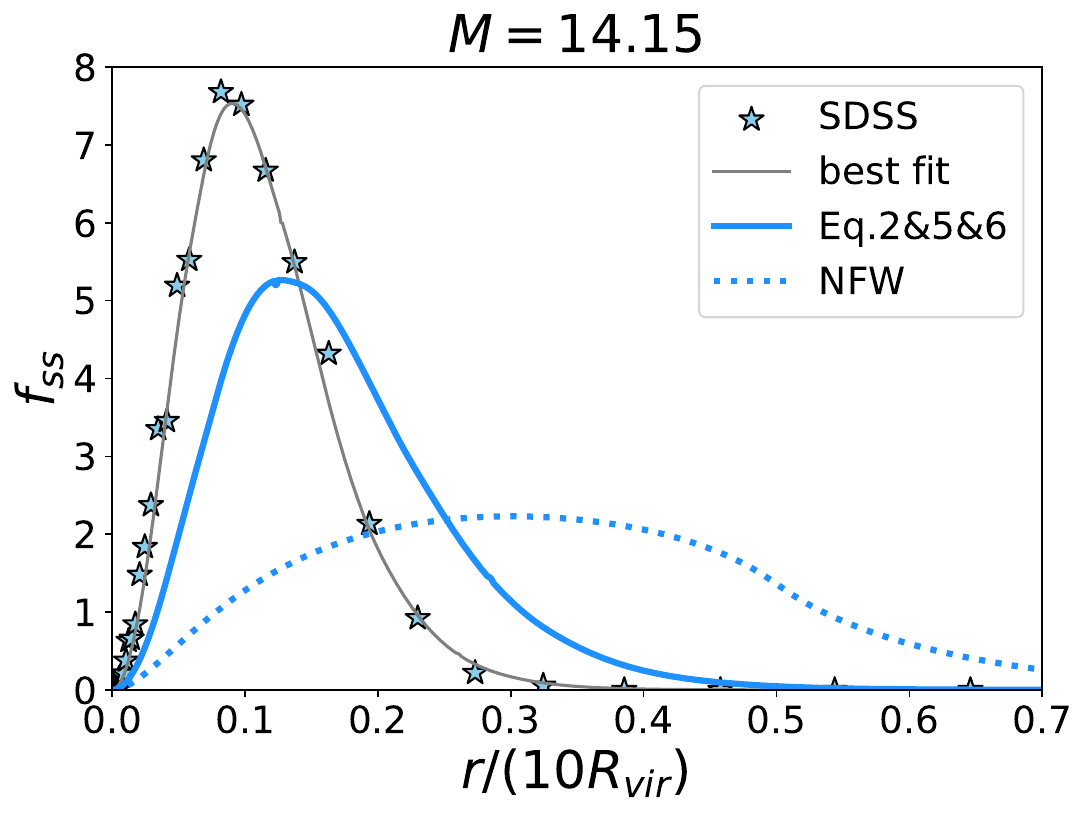}

 \caption{Comparing the $f_{cs}$ and $f_{ss}$  of SDSS galaxies to those of \ds~and NFW profile, in an example halo mass bin $M=14.15$. 
} 
 \label{sdssFcsY}
\end{figure}

In Fig.~\ref{sdssFcsY}, the blue filled pentagrams in the top  panel show the measured $f_{cs}$ in an example halo mass bin $M=14.15$. 
The grey curve is the model Eq.~\ref{newprof} fitting $\alpha$ and $\beta$ to the blue filled circles. The dashed-blue curve is the prediction from the NFW profile, which does not agree with the measured points.   
The solid-blue curve is the prediction of the new model for  $M=14.15$ from Eq.~\ref{newprof},  using the mass-dependent parameter values for $\alpha$ (Eq.~\ref{Am}) and $\beta$ (Eq.~\ref{Bm}) (i.e. parameter values inferred from \ds). Though it is slightly broader than the measurements, it is much better than the NFW profile (blue dashed curve) in matching the data (blue filled pentagrams). 
The discrepancy of \ds~ (the solid-blue curves) and SDSS (the blue filled pentagrams) could be caused by subhalo-finder issues in Millennium.  The closer to the halo centre a subhalo is, the more likely a subhalo finder will fail to find it.

The blue filled pentagrams in the bottom  panel of Fig.~\ref{sdssFcsY} show the measured $f_{ss}$ of SDSS galaxies, the grey curve is calculated the same as before, using the best fit values of $\alpha$ and $\beta$ ( i.e.  the red stars in Fig.~\ref{fitAB}) into Eq.~\ref{convnew}.  The dashed-blue curve is the prediction from the NFW profile  which is poorly agree with the measurement.  
The solid-blue curve is a prediction for $M=14.15$  from Eq.~\ref{convnew}, using the mass-dependent parameter values for $\alpha$ (Eq.~\ref{Am}) and $\beta$ (Eq.~\ref{Bm}).
It is  broader than the measurements, however it is a much better fit than the NFW profile to the blue filled pentagrams. 

{The minimum separation of the spectroscopic fibers  of the telescope is 55 arcseconds \citep{Tempel2014}, as for any smaller separation the fibers will collide. The distance limit of the sample of SDSS galaxies used in this paper is 330 Mpc $h^{-1}$ ($z<0.11$). Therefore, while the fiber collsions may lead to some incompleteness, this will bias the measurements for a separation of $r< 0.088$ Mpc $h^{-1}$ at most. 
The halo mass range of the SDSS data used in this paper is $M\in[13.5,~14.8]$. Correspondingly, the range of virial radius is $R_{vir}\in[0.53,~1.35]$ Mpc $h^{-1}$. Therefore, using the SDSS data, the measurements of $f_{cs}$ (and $f_{ss}$) in $x<$0.0187  may not be reliable.  However, this small-scale incompleteness is not expected to be a major problem. As illustrated by Fig.\ref{sdssFcsY}, these measurements are located in the left-corner of the figure, and so they will not affect our fit too much.}

 Under the NFW profile, the satellite galaxies are much more close to the halo center, therefore, the gravitational  potential and redshift calculated from the NFW profile are lower than the measurements, this will yeild incorrect measurements of correlation function odd-multi-pole, further more, bias the estimations of the cosmological parameters.

\subsection{Galaxy two-point correlation function of SDSS galaxies}

\begin{figure} 
\centering
 \includegraphics[width=\columnwidth]{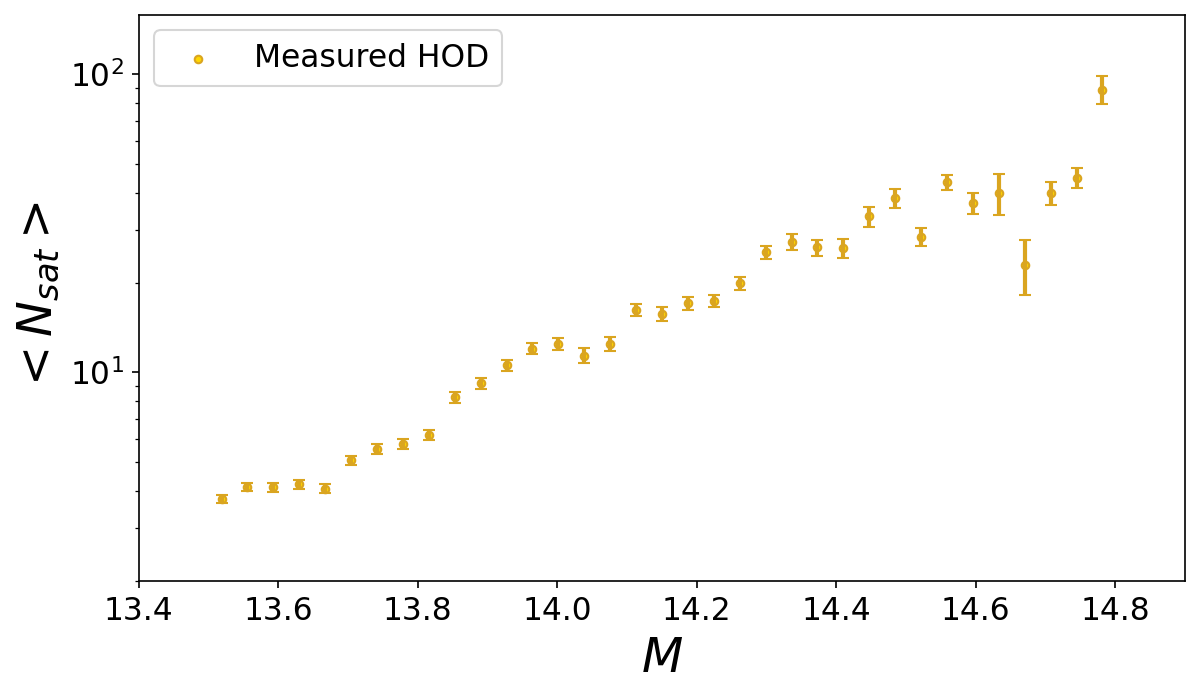}
  \includegraphics[width=\columnwidth]{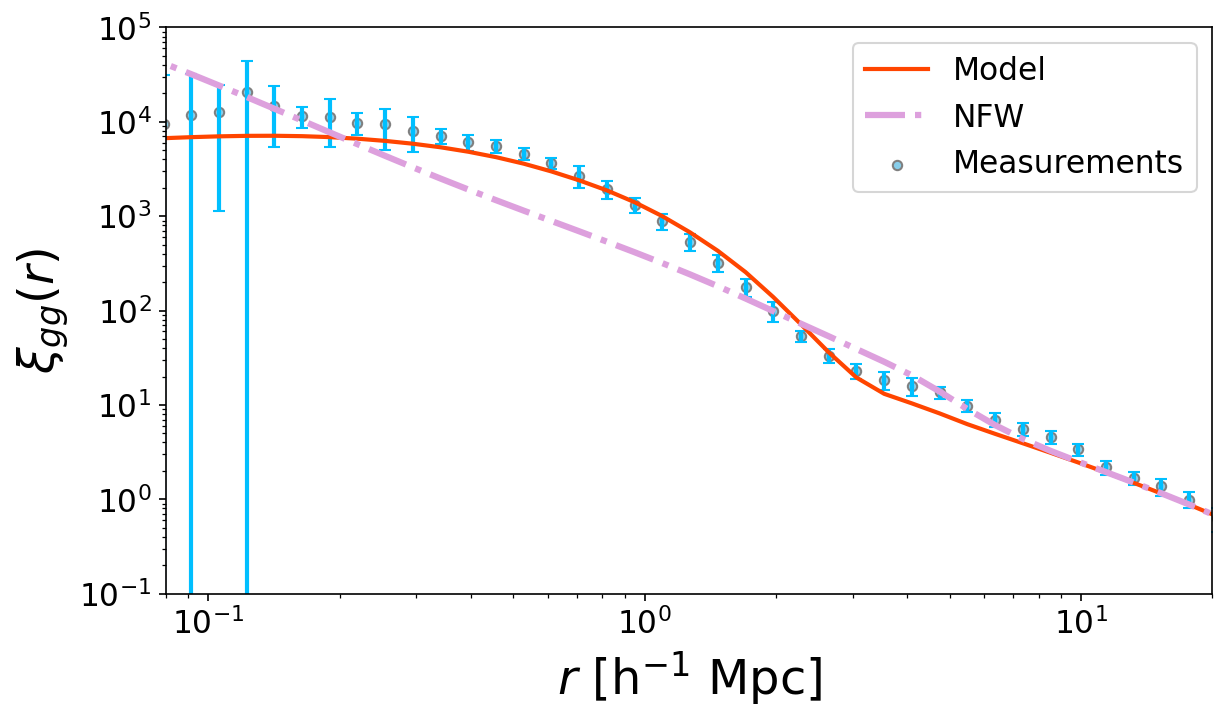} 
 \caption{The top panels show the HOD of satellite galaxies measured from the SDSS galaxies. The bottom panel shows the galaxy two-point correlation functions $\xi_{gg}(r)$. The blue dots are the measurements from SDSS galaxies. The red solid curve is the model $\xi_{gg}(r)$ inferred from the measured HOD in the top panel and the halo density profile for galaxies Eq.~\ref{newprof}, \ref{Am} and \ref{Bm}.  The pink dash-dotted curve is the model $\xi_{gg}(r)$ inferred from  the measured HOD in the top panel and the NFW profile. 
} 
 \label{sdssxi}
\end{figure}

The top panel of Fig.~\ref{sdssxi} shows the HOD  of satellite galaxies $N_{\rm sat}$ measured from the SDSS galaxies.  
After plugging $N_{\rm sat}$ and the galaxy number density profile model of haloes (Eqs~\ref{newprof}, \ref{Am}, \ref{Bm})  into  Eq.~\ref{realxigg}, we show our model correlation function in the red solid curve of the bottom panel. We also plug $N_{\rm sat}$ and the NFW profile model into Eq.~\ref{realxigg} to calculate the model correlation function, as shown in the pink dash-dotted curve of the bottom panel. The blue circles are the measurements from the SDSS galaxies.

Again, the correlation function predicted by
our new galaxy number density profile of haloes
is in much better agreement with the measurements (blue dots) than the NFW case. 
The discrepancy between our model and the data reflects the difference of the best-fitting $\alpha$ and $\beta$ values between \ds~and SDSS. 
The SDSS galaxies tend to be more concentrated to the halo center, therefore, the blue dots  for small scales $r \leq 0.7\,h^{-1}$ Mpc are higher than the red curve, while the blue dots for intermediate scales $r\in[0.7,~3] h^{-1}$ Mpc are lower than the red curve. 

{As mentioned in the Section \ref{sec:5.1}, due to the incompleteness from the fiber collision,   
 the correlation function measurements below 0.088 Mpc $h^{-1}$ are not reliable. However, our analyse is mainly for $r>0.1$ Mpc $h^{-1}$ in Fig.\ref{sdssxi}. Therefore, the incompleteness will not affect our findings.
}

\section{Conclusion} \label{sec:conc}
 
In this paper, we have presented a new galaxy number density profile model of haloes (Eq.~\ref{newprof}). The functional form of this model is determined using the \ds~semi-analytic model of galaxy formation. There are two free parameters ($\alpha,~\beta$) in the model. Based on the measurements from the  \ds~simulation, we find that  $\alpha$ and $\beta$ are relatively simple functions of halo mass $M$. We can use a linear function  Eq. \ref{Am} to model the relation between $\alpha$ and  $M$, and use a exponential function Eq. \ref{Bm} to model the relation between $\beta$ and  $M$. We also find that our  galaxy number density profile model of haloes can correctly reproduce the galaxy two-point correlation function of the \ds~galaxies. 

We derived the analytic expressions for the circular velocity and gravitational potential energy (see Eqs. \ref{criclV} and \ref{gwphi}) for the new  model Eq.~\ref{newprof}. 

We also used the SDSS DR10 galaxy group catalogue to validate the galaxy number density profile of haloes and find that the functional form of our model (Eqs. \ref{newprof}, \ref{Am} and \ref{Bm}) is flexible enough to fit the data, although the best fit values of ($\alpha,~\beta$) are somewhat larger than the fit results from \ds. The galaxies in the SDSS catalogue is more concentrated to the halo center compared to the galaxies of \ds. The galaxy two-point correlation function predicted from our galaxy number density profile model of haloes  is comparable to the measurement from SDSS galaxies.

The NFW density profile fails to model the radial distribution of satellite galaxies in their host halo for both the \ds ~simulation and SDSS catalogue. And the NFW profile also fails to predict the two-point correlation functions for both the \ds ~simulation and SDSS catalogue in small separation scales.

There are many extensions to this work that can be pursued in the future,  including using larger galaxy group catalogues and simulations to further constrain $\alpha$ and $\beta$ at higher and lower halo masses, exploring the galaxy number density profile of haloes conditional on \HI~mass, testing our galaxy number density profile of haloes using emission line galaxies, and investigating how the galaxy number density profile of haloes affects the measurements of odd-multipoles of the correlation function.

\acknowledgments

FQ and DP are supported by the project \begin{CJK}{UTF8}{mj}우주거대구조를 이용한 암흑우주 연구\end{CJK}, funded by the Ministry of Science. CH is supported by the Australian Government through the Australian Research Council’s Laureate Fellowship and Discovery Project funding schemes (projects FL180100168 and DP20220101395).
ARHS is funded through the Jim Buckee Fellowship at ICRAR-UWA.

%



\software{ \textsc{Nbodykit} \citep{Nbodykit2018},  
      \textsc{ChainConsumer} \citep{ChainConsumer},
           \textsc{emcee} \citep{Foreman-Mackey2013}, 
          \textsc{Dark Sage} \citep{Stevens2017code},
          \textsc{Scipy} \citep{Virtanen2020},       \textsc{Corrfunc}  \citep{Corrfunc2020},  
          \textsc{Matplotlib} \citep{Hunter2007},
           \textsc{CAMB} \citep{Lewis:1999bs}.                   
}

\appendix

\section{NFW Halo Density Profile}\label{sec:nfw}

The Navarro--Frenk--White (NFW) profile \citep{NFW1996,Navarro1997}
\be\label{NFWden}
\rho_m(R)=\frac{\rho_s}{(R/R_s)(1+R/R_s)^2}~,
\ee 
where
\be \label{NFWdencdcd}
\rho_s=\frac{\rho_0\Delta}{3}\frac{c^3}{\ln(1+c)-c/(1+c)}~,
\ee  
The $f_{ss}(x)$ for the NFW profile is given in the Appendix A of \citet{Zheng2007}, the expression is lengthy so not repeated here.  
The Fourier transformation of the NFW profile is  expressed as \citep{Cooray2002}
\be \label{ygnfw}
\begin{split}
y(k)&=\frac{1}{\ln(1+c)-c/(1+c)}  \times \bigg \{ -\frac{\sin(ckR_s)}{(1+c)kR_s} \\
& +\sin(kR_s)\left[SI([1+c]kR_s)-SI(kR_s)  \right]   \\
&  +\cos(kR_s)\left[CI([1+c]kR_s)-CI(kR_s)  \right] 	\bigg \} ~,
\end{split}
\ee 
where
\be 
SI(x) \equiv \int_0^x\frac{\sin t}{t}dt~,
CI(x) \equiv -\int_x^{\infty}\frac{\cos t}{t}dt~,
\ee 
are the Trigonometric integrals.  
The mass interior
to  $r$ is given by \citep{Sheth2001a}
\be  
m(<r)=\frac{m_{\rm vir}}{R_{\rm vir}}\frac{r}{a[\ln(1+1/a)-1/(1+a)]}\left[\frac{\ln(1+x)}{x}-\frac{1}{1+x}\right]
\ee  
The circular velocity is \citep{Sheth2001a}
\be 
v^2_c= \frac{Gm_{\rm vir}}{R_{\rm vir}}\frac{1}{a[\ln(1+1/a)-1/(1+a)]}\left[\frac{\ln(1+x)}{x}-\frac{1}{1+x}\right]
\ee 
where $x=r/a$ and $a=1/c$. The gravitational potential energy is given by \citep{Sheth2001a}
\be  
\phi(r)= -\frac{Gm_{\rm vir}}{R_{\rm vir}}\frac{1}{a[\ln(1+1/a)-1/(1+a)]}\left[\frac{\ln(1+x)}{x}-\frac{a}{1+a}\right]
\ee

\section{HOD parameters}\label{sec:hodp}

\begin{figure} 
\centering
  \includegraphics[width=\columnwidth]{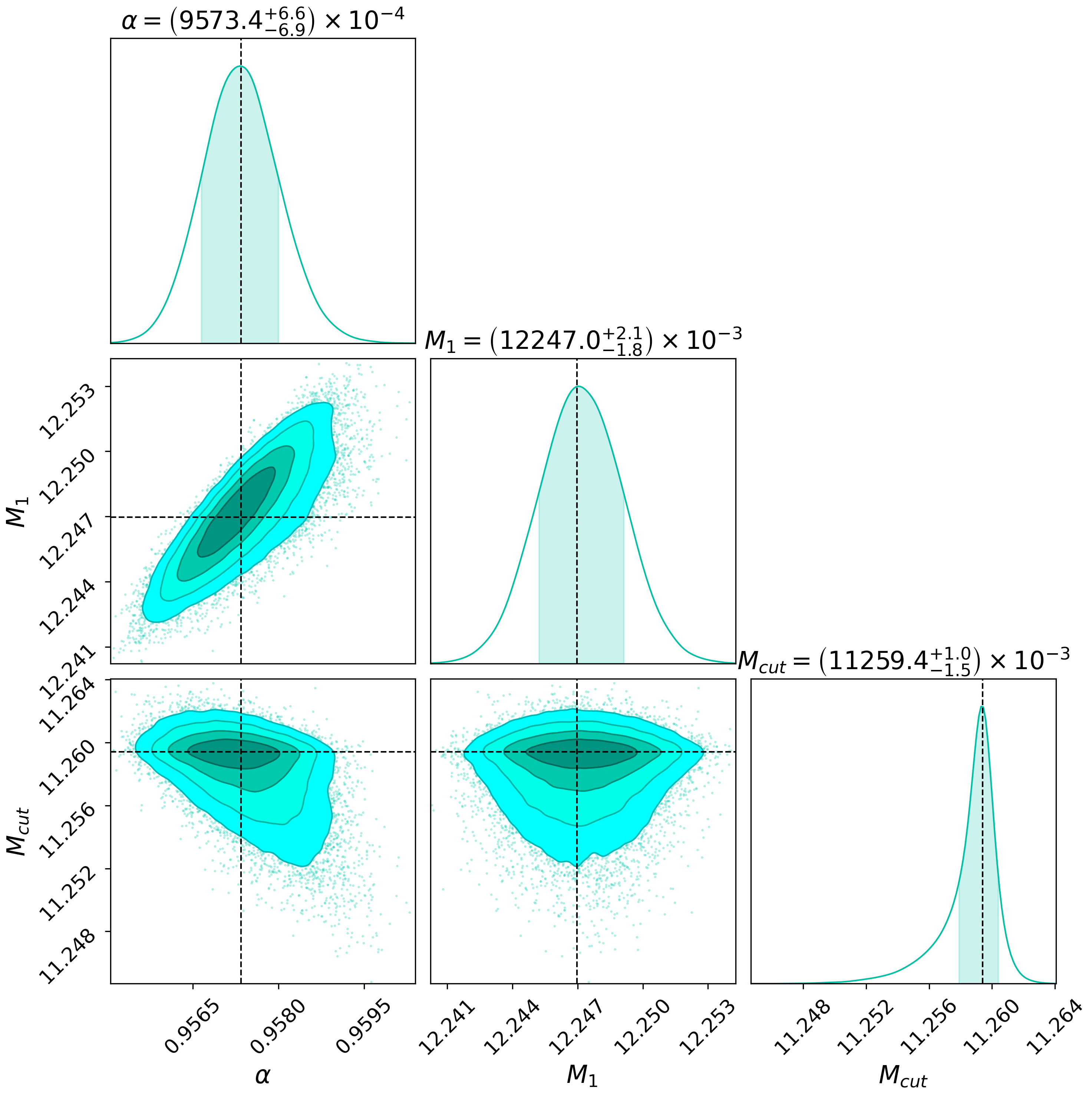}
 \caption{The fit results of the HOD parameters. The histograms display the distribution of out MCMC samples. The shaded areas in the 1D histograms indicate the 68\% confidence level, while the 2D contours indicate the 1, 1.5, 2 and 2.5$\sigma$ regions.
} 
 \label{xigghod222}
\end{figure}

{Corresponding to Fig.\ref{xigghod}, 
the fit results of the HOD parameters in Eq.\ref{satHODDS}   is shown in Fig.\ref{xigghod222}. }
 
\bibliography{FQinRef}{}
\bibliographystyle{aasjournal}



\end{document}